\documentclass[useAMS,usenatbib]{mn2e}
\usepackage{amsmath}
\usepackage{txfonts}
\usepackage{epsfig}
\eqsecnum

\newcommand\msun{{{{\rm M_\odot}}}}
\def\spose#1{\hbox to 0pt{#1\hss}}
\def\lta{\mathrel{\spose{\lower 3pt\hbox{$\sim$}}
    \raise 2.0pt\hbox{$<$}}}
\def\gta{\mathrel{\spose{\lower 3pt\hbox{$\sim$}}
    \raise 2.0pt\hbox{$>$}}}

\def\zl{z_{\rm L}}

\def\rpetro{{r_{\rm pet}}}
\def\rpsf{{r_{\rm psf}}}

\title[Two New Gravitational Lenses]{Two New Large Separation Gravitational Lenses from SDSS}

\author[Belokurov et al.]{V. Belokurov $^1$, N.W. Evans $^1$,
  P.C. Hewett $^1$, A. Moiseev$^2$, R.G.~McMahon$^1$, \newauthor
  S.F. Sanchez$^3$ and L.J.~King$^1$ \\
$^1$ Institute of Astronomy, University of Cambridge, Madingley Road, Cambridge,CB3 0HA, United Kingdom\\
$^2$ Special Astrophysical Observatory, Nizhniy Arkhyz,
Karachaevo-Cherkessiya, Russia\\
$^3$ Centro Astronomico Hispano Aleman de Calar Alto
  (CSIC-MPIA), E4004 Almería, Spain\\
}

\begin{document} 

\maketitle

\begin{abstract}
  We present discovery images, together with follow-up imaging and
  spectroscopy, of two large separation gravitational lenses found by
  our survey for wide arcs (the CASSOWARY). The survey exploits the
  multicolour photometry of the Sloan Digital Sky Survey to find
  multiple blue components around red galaxies. CASSOWARY~2 (or ``the
  Cheshire Cat'') is composed of two massive early-type galaxies at $z
  = 0.426$ and $0.432$ respectively lensing two background sources,
  the first a star-forming galaxy at $z = 0.97$ and the second a high
  redshift galaxy ($z> 1.4$). There are at least three images of the
  former source and probably four or more of the latter, arranged in two
  giant arcs. The mass enclosed within the larger arc of radius $\sim
  11^{\prime\prime}$ is $\sim 33 \times 10^{12}\ \msun$. CASSOWARY~3
  comprises an arc of three bright images of a $z = 0.725$ source,
  lensed by a foreground elliptical at $z = 0.274$. The radius of the
  arc is $\sim 4^{\prime\prime}$ and the enclosed mass is $\sim 2.5
  \times 10^{12}\ \msun$.  Together with earlier discoveries like the
  Cosmic Horseshoe and the 8 O'Clock Arc, these new systems, with
  separations intermediate between the arcsecond separation lenses of 
  typical strong galaxy lensing and arcminute separation cluster lenses,
  probe the very high end of the galaxy mass function.
\end{abstract}


\section{Introduction}

Very recently, a number of large separation gravitational lenses have
been found in data from the Sloan Digital Sky Survey (SDSS). The systems
include: the 8 O'clock Arc, which is a Lyman Break galaxy lensed into
three images merging into an extended arc~\citep{Al07}, the Cosmic
Horseshoe, which is a star-forming galaxy lensed into an almost
complete Einstein ring of diameter $10^{\prime\prime}$ ~\citep{Be07},
and the strongly lensed post-starburst galaxy of~\citet{Sh08}.

These discoveries prompted us to instigate the {\it The CAmbridge
Sloan Survey Of Wide ARcs in the skY}
(CASSOWARY~\footnote{http://www.ast.cam.ac.uk/research/cassowary/}). The
aim is to carry out a systematic search for wide separation
gravitational lens systems, looking for multiple, blue companions
around massive ellipticals in the SDSS photometric catalogue. Typically,
the target systems correspond to lensing of $z \gtrsim 0.5$
star-forming galaxies by luminous red galaxies and brightest cluster
galaxies.  Here, we present a description of the search strategy, together
with details of two new gravitational lens systems, for which we have
obtained follow-up on a number of telescopes, including the 6 m at the
{\it Special Astrophysical Observatory} (SAO), the 3.5 m at {\it Calar
Alto}, and the 2.5 m {\it Isaac Newton} (INT) and the 4.2 m {\it
William Herschel Telescopes} (WHT) at La Palma. The accompanying
CASSOWARY webpage lists 23 likely candidates, many still awaiting
follow-up.

Previous search strategies typically target smaller separation lenses, in which
the images are unresolved by SDSS.  For example, \citet{In03b} and \citet{Jo03}
searched through spectroscopically identified quasars, looking for evidence for
extended sources corresponding to unresolved, multiple images, whilst
\citet{Jo03}, \citet{Bo04} and \citet{Wi05} used the spectroscopic database to
look for emission lines of high redshift objects within the spectrum of lower
redshift early-type galaxies.

Although wide separation lenses ($\gta 3^{\prime\prime}$) are
comparatively unexplored, they are interesting for a number of
reasons.  First, wide separation lenses are a probe of the high-mass
end of the galaxy mass function. The Einstein radius is typically a
few effective radii where the matter distribution is dominated by dark
matter. Second, the source is often highly magnified and thus provides
us with a sample of the brightest galaxies known at high redshifts, such
as the Cosmic Horseshoe or CASSOWARY~1 \citep{Be07, Dy08}.
Third, the modelling of such systems is relatively clean.  The
positioning of the images well outside the effective radius of the
lens means that their properties, particularly their brightnesses, can
be measured with high accuracy.  Fourth, the frequency of large
separation lenses provides constraints on models of structure
formation. For example, fossil groups, in which bright galaxies have
merged via dynamical friction to leave a single very massive object,
are amongst the lenses targeted by the CASSOWARY search. Finally, the
unusual morphologies of some wide separation lenses are also
fascinating from the perspective of the theory of gravitational
lensing (Shin \& Evans 2008, Werner, An \& Evans 2008).

The paper is arranged as follows. Section 2 presents our general
methodology, whilst Sections 3 and 4 present our two new gravitational
lenses in turn -- CASSOWARY 2 and CASSOWARY 3 (henceforth CSWA~2 and
CSWA~3) -- describing our follow-up data, our modelling and
predictions. They are the second and third of the CASSOWARY lenses,
following the Cosmic Horseshoe, CSWA~1.  Finally, Section 5 summarizes our
conclusions and future prospects.

\begin{figure*}
\begin{center}
\includegraphics[height=12cm]{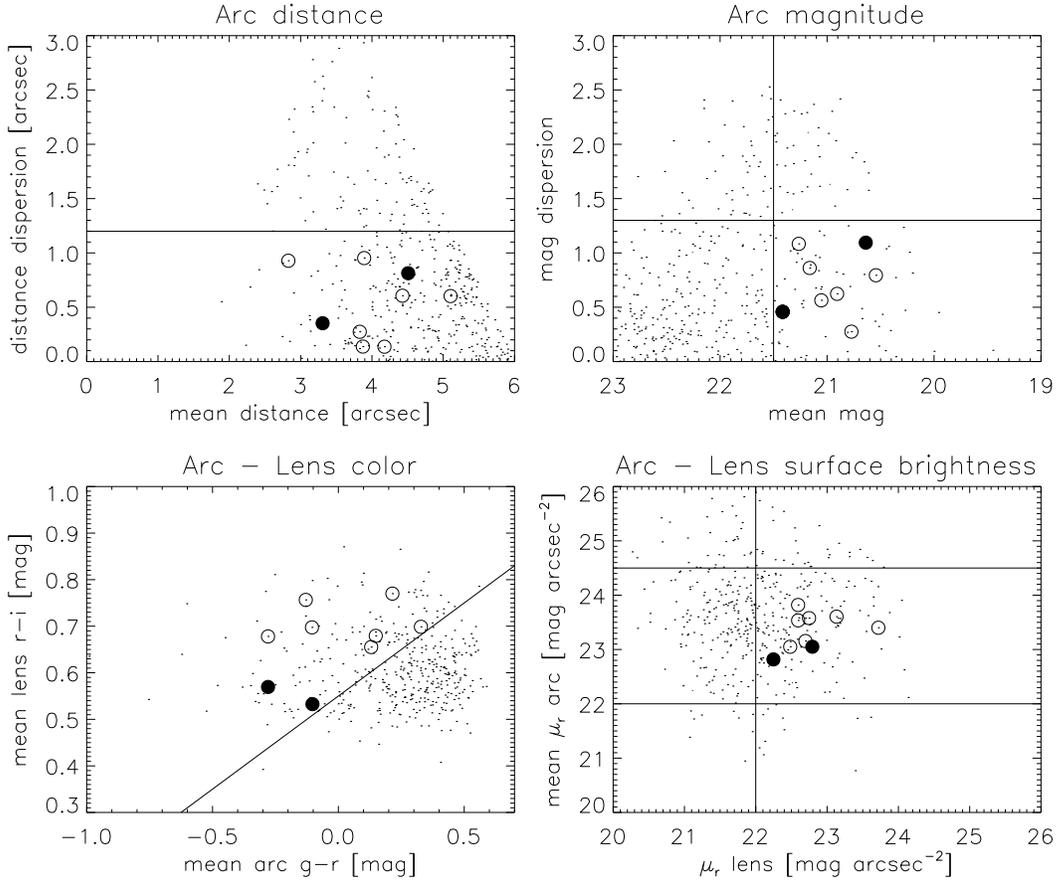}
\caption{\label{fig:algorithm} Illustration of the cuts to select our
  candidates. Dots represent arc
  candidates from the initial large sample, 
  circles are objects selected by the cuts on distance,
  magnitude, colour and surface brightness given in eqn~(2.2).  Filled
  circles are the two known systems (the 8 O'Clock Arc and the Cosmic
  Horseshoe), whilst there are seven open circles, corresponding to six
  distinct candidates.}
\end{center}
\end{figure*}
\begin{figure*}
\begin{center}
\includegraphics[height=8cm]{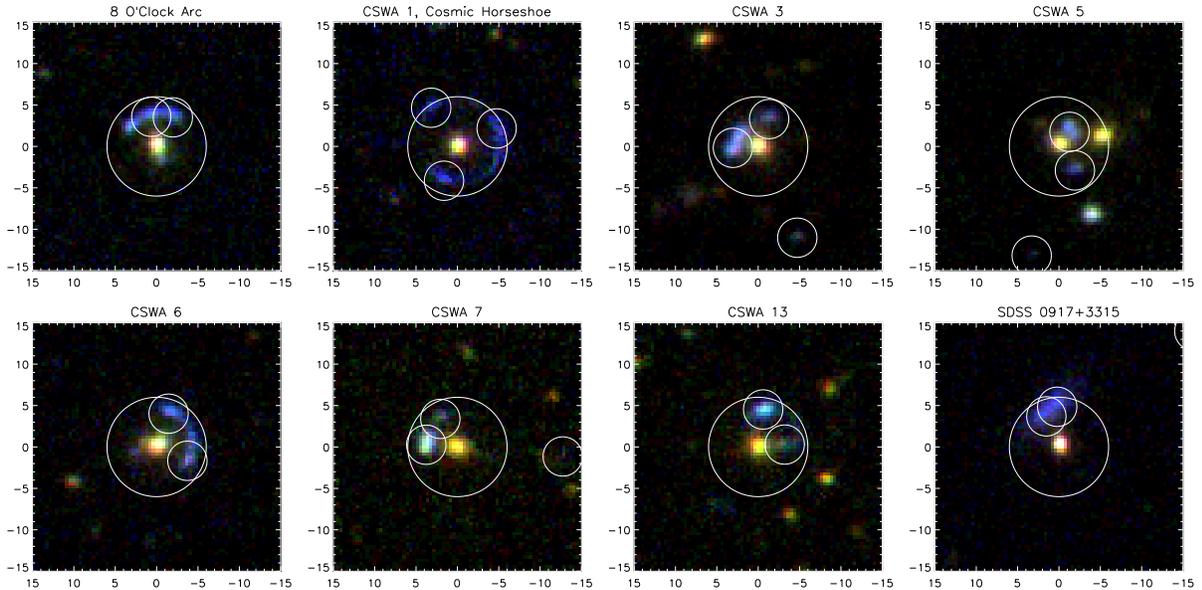}
\caption{\label{fig:cutouts} SDSS $g,r,i$ composite images,
  $30^{\prime\prime}$ on a side, in conventional astronomical orientation, 
  of the eight
  objects selected by the cuts. The large circle marks the
  $6^{\prime\prime}$ search radius around the primary, the small circles
  show the locations of blue companions selected by the SQL query.
  The candidate name is given above each panel.}
\end{center}
\end{figure*}

\section{Method}

\subsection{Search Strategy}

At its simplest, our algorithm searches for blue companions to
luminous red galaxies. We have already used a precursor of the
algorithm to identify the Cosmic Horseshoe~\citep{Be07}, and similar
methods have been used recently by others \citep{Ku07,Sh08}.

Our selection procedure has two parts. First, we carry out a broad
search for all massive ellipticals with at least one blue companion.
Secondly, we run a number of more
targetted searches for arcs of different sizes and brightnesses on
the initial catalogue of candidates.

To obtain the initial sample, we use a variant of the criteria of \cite{Ei01}
to identify luminous red galaxies (LRGs) in SDSS's {\tt Galaxy} table.
Specifically, we require
\begin{eqnarray}
&&\rpetro < 14 + c_\parallel /0.3,\qquad \rpetro < 19.5,\qquad
|c_\perp | < 0.2,\nonumber \\ 
&&\mu_r < 24.2, \qquad \rpsf - r > 0.3.
\end{eqnarray}
These quantities are all defined in \citet{Ei01}, but we use the Data
Release 6 (DR6) photometric solution rather than the Early Data
Release magnitudes. Our cuts are slightly changed from \citet{Ei01}
to include fainter and somewhat bluer objects than the classical LRGs, as we
are primarily interested in early-type massive galaxies as opposed to
LRGs themselves.  Following~\cite{Ei01}, we only keep galaxies
with $g-r < 2.5$ and $r-i < 1.5$. We also use an $i > 17.0$ cut to eliminate
nearby late-type galaxies that are resolved by SDSS
into separate bulge, disk and spiral arm components.

We then select objects from SDSS's {\tt Neighbors} table with $g-r
<0.6$ and $19.0 < r < 23.5$ within $30^{\prime\prime}$ radius. The SQL
query was based on one originally devised by R. Lupton~\footnote{
  http://cas.sdss.org/dr6/en/help/docs/realquery.asp\#nbrrun}. Some
standard photometry flags are also used to remove artifacts,
specifically the first two lines from the SDSS sample SQL query
``Clean Photometry with flags'' for the {\tt Galaxy} view~\footnote{
  http://cas.sdss.org/dr6/en/help/docs/realquery.asp\#flags}

The second stage begins with the calculation of the total number of
arc candidates and the mean values of their distance, apparent
magnitude, colour and surface brightness for each potential lens
galaxy.  Selection cuts are then applied to eliminate false positives 
in the form of galaxy groups, mergers, tidally induced structures 
and data artifacts.  In their simplest form, they are as follows:
\begin{eqnarray}
&&D < 6^{\prime\prime},\qquad \sigma_D < 1.2^{\prime\prime}, \\
&&\langle r_{\rm arc} \rangle  < 21.5, \qquad \sigma_r  < 1.3,\\
\label{eq:algorithm}
&&(r-i)_{\rm lens}  > 0.4 \langle (g-r)_{\rm arc} \rangle + 0.55,\\ 
&&\mu_{\rm lens} > 22,\qquad 22 < \mu_{\rm arc} < 24.5.
\end{eqnarray}
Thus, the search requires the presence of at least two arc candidates, that is,
blue faint companions lying within $D =6^{\prime\prime}$ radius from
the center of the lens galaxy.  If the lens is a single
galaxy with modest shear, the images should be all at about the same
distance from the primary and the distance dispersion $\sigma_D$
should be small. The arcs are required to be bright (on average
brighter than 21.5 mag) and to have at least two components of
comparable magnitude ($\sigma_r < 1.0$ mag).  Eqn~(\ref{eq:algorithm})
ensures that at each lens colour, the arcs are bluer than the typical
physical companions that can occur. The final cut restricts the allowed 
range of surface brightness for the arcs $\mu_{\rm arc}$.
Brighter objects are typically foreground galaxies and the
fainter objects are typically tidal arms and bridges common in interacting
galaxies.

The cuts are illustrated in the four panels of
Figure~\ref{fig:algorithm}.  The filled and open circles show the
locations of nine objects selected by our algorithm. One lens is actually
double-counted, as it is a binary, so there are really only eight distinct
cases.  The two solid black circles are the previously known lenses:
the 8 o'clock Arc~\citep{Al07} and the Cosmic Horseshoe, or CSWA~1
\citep{Be07}. It is clear that the new candidates, shown as black
circles, look similar to the known systems, but are on average not as
blue and do not have as high surface brightness. The SDSS cut-outs of
the eight systems selected by the algorithm are shown in
Figure~\ref{fig:cutouts}.  Although they all look excellent
gravitational lens candidates, they require follow-up before their
nature can be established beyond doubt.

\begin{figure}
\begin{center}
\includegraphics[height=6cm]{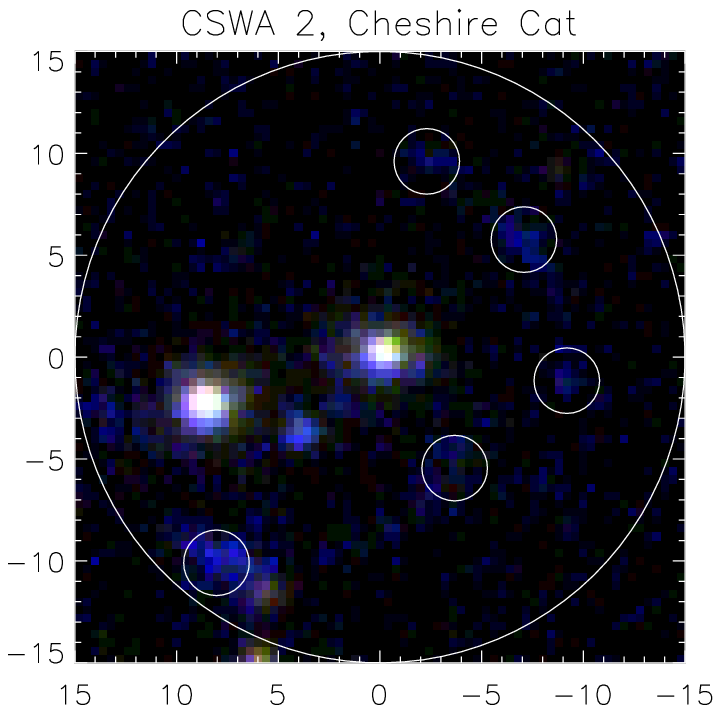}
\caption{\label{fig:cutouts2} SDSS $g,r,i$ composite image,
  $30^{\prime\prime}$ on a side, in conventional astronomical orientation
  of CSWA~2,
  or the ``Cheshire Cat''. The large circle marks the search area around
  the primary, the small circles show locations of blue companions
  selected by our cuts. }
\end{center}
\end{figure}

Relaxing the cuts described in eq~(\ref{eq:algorithm}), to allow
fainter, less circular arcs at larger radii from the lens, yields many
additional candidates at the expense of including a larger fraction of
false positives.  One particularly interesting system is shown in the
SDSS cut-out in Figure~\ref{fig:cutouts2}.

The results of our searches are summarised on the CASSOWARY website,
which includes 20 gravitational lens candidates.  For
two of the systems, CSWA~2 and CSWA~3, we have obtained deeper imaging and
spectroscopy, confirming the gravitational lensing hypothesis. We now
discuss the properties of CSWA~2 or the ``Cheshire Cat'' and CSWA~3 in 
some detail.

\subsection{Overview of the Properties of CSWA~2 and CSWA~3}

For CSWA~2, deeper imaging observations were carried out on 2007 May
12/13 with the Wide Field Camera (WFC) of the {\it Isaac Newton
  Telescope} (INT).  The exposure times were 600\,s in each of the
$U, g$ and $i$ filters -- which are similar to the SDSS
filters.  The measured seeing (FWHM) on the images
($0.33^{\prime\prime}$ pixels) was $1.30^{\prime\prime}$,
$1.26^{\prime\prime}$ and $1.21^{\prime\prime}$ in $U, g$ and $i$
respectively. The INT data are roughly a magnitude deeper than the
SDSS data and were reduced using the pipeline toolkit of
~\citet{Ir01}. Although less extensive, we also obtained some
additional $R$ band imaging of CSWA~3 using the {\it Special
  Astrophysical Observatory} (SAO) on 2008 March 28/29.  BTA~6-m
telescope at the SAO in seeing of $1.1^{\prime\prime}$.

\begin{figure*}
\begin{center}
\includegraphics[height=4cm]{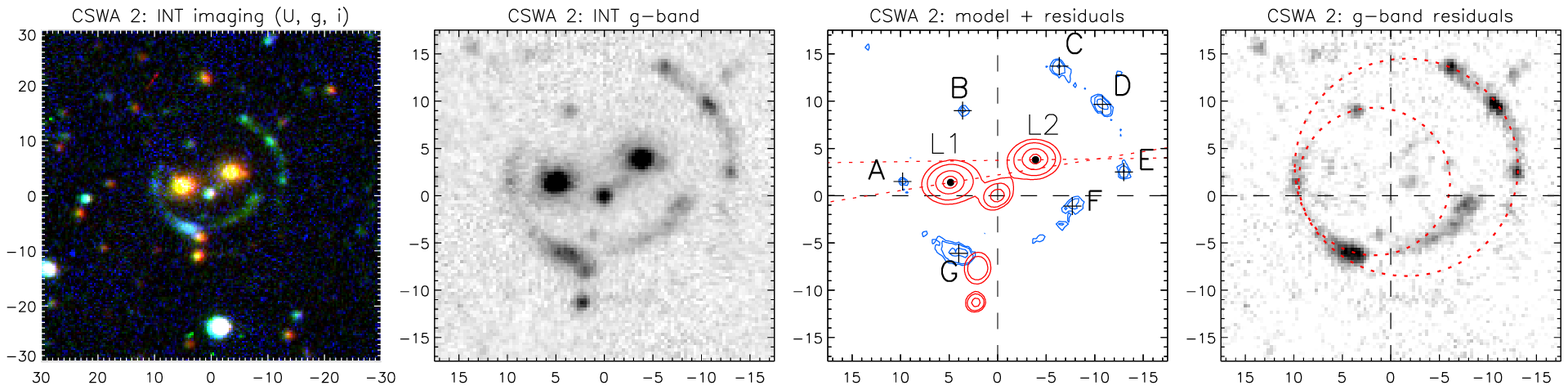}
\includegraphics[height=4cm]{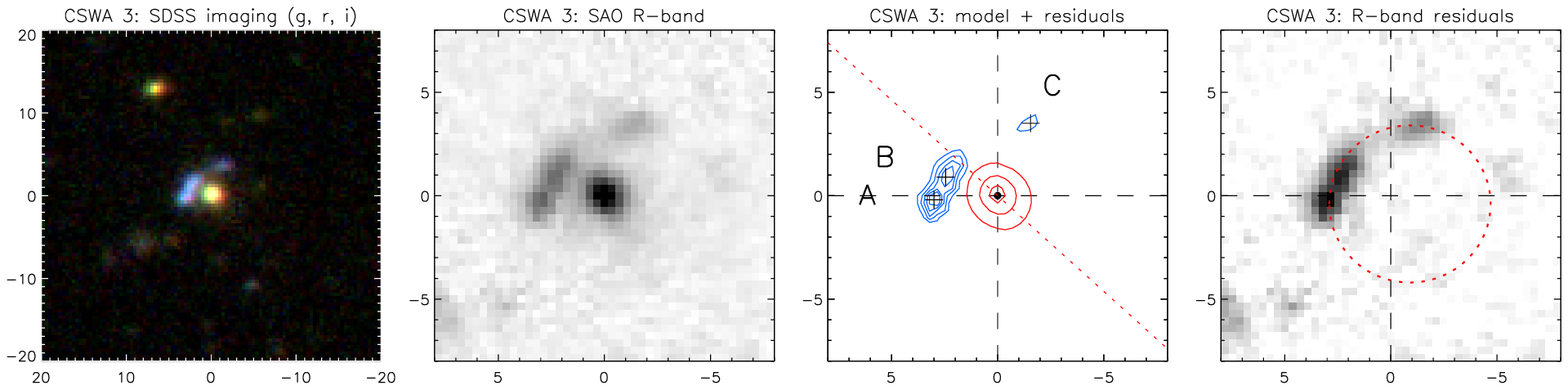}
\caption{\label{fig:cassdata} Left: Best available RGB composite
  (either SDSS or follow-up) for CASSOWARY lenses. Middle: Zoom-in of
  $g$ ($R$) band image for CSWA~2 (CSWA~3). Right: GALFIT model to the
  lens population (red) and residuals after model subtraction (blue).
  All images are in conventional astronomical orientation with the axes
  indicating the scale in arcseconds.}
\end{center}
\end{figure*}

In Figure~\ref{fig:cassdata}, we present RGB-composites made from
$U,g,i$ in the case of CSWA~2 and $g,r,i$ in the case of CSWA~3. A
zoom-in of the $g$ (or $R$) band image for CSWA~2 (or CSWA~3) is shown
in the second column.  In the case of CSWA~2, the arcs are more
extended in Figure~\ref{fig:cassdata} than in the SDSS image, with at
least seven resolved bright knots visible. Moreover, the arc
connecting components A and B together with the knot G are bluer than
the arcs connecting C, D, E and F (see the top panel in the third
column for the component key). The colour difference hints at the
possibility of two sources at different redshifts in CSWA~2.

For both CSWA~2 and CSWA~3, we masked the arcs in the $g$ band to
identify the deflector population, which is modelled using {\tt
  GALFIT}~\footnote{http://users.ociw.edu/peng/work/galfit/galfit.html}~\citep{Pe02}. We
use a field star to define the PSF and fix the sky levels from
aperture photometry of stars around each lens. Models of the
foreground galaxies (and occasional star) are shown in red in the
third column of Figure~\ref{fig:cassdata}, whilst the residuals --
that is, primarily the arcs -- are shown in blue. The red dotted lines
show the major axes of the lens galaxies. The fourth panel shows the
greyscale image with the foreground population subtracted, together
with the best-fitting ellipses to the arcs. The ellipses are used to 
estimate the radius of curvature of the arcs.

Table~\ref{tab:struct} gives the basic observational data for the
CSWA~2 and CSWA~3 lenses, together with the Cosmic Horseshoe from
\cite{Be07} for comparison. All magnitudes are given in the
$g$ and $i$ bands on the SDSS AB system. The $i$ band images of the
lensing galaxies have the highest signal-to-noise ratio, and so they
are used to measure the effective radii using a standard de
Vaucouleurs profile fit. Anticipating the results from the next two sections,
our follow-up spectroscopy is used to determine the lens redshift for
CSWA~2 and CSWA~3, whilst the mass estimates come from our modelling.

\begin{figure*}
\begin{center}
\includegraphics[height=5cm]{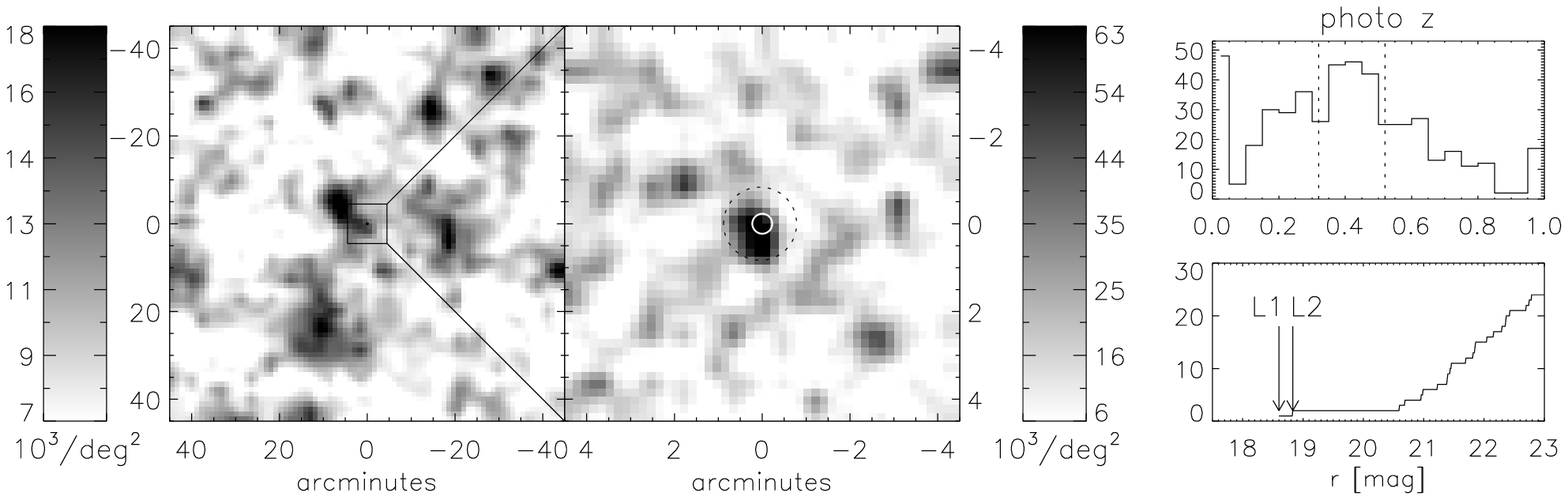}
\includegraphics[height=5cm]{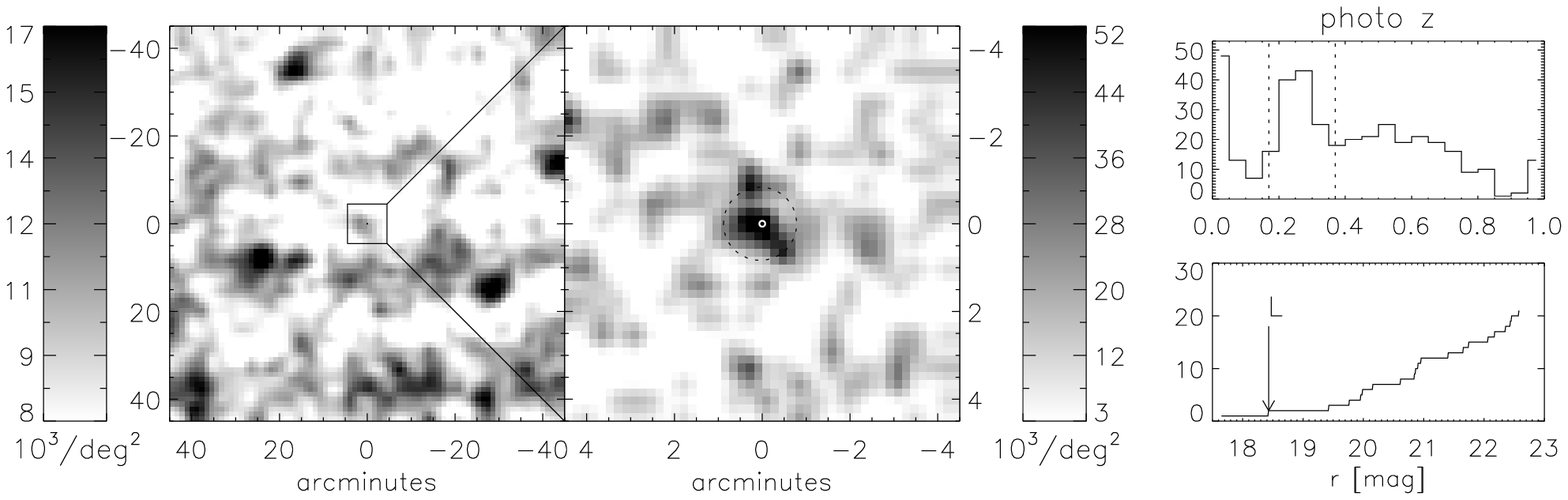}
\caption{\label{fig:cassenviron} Surface density of galaxies in the vicinity
  of each lens with corresponding redshift range marked.  Left: Large
  scale structure. Middle: Zoom-in on the lens marked by white ring,
  shown to scale. Right: Redshift distribution (upper panel) for all
  galaxies in the dashed circle in the middle panel. For galaxies in
  the range marked, we build the $r$-band cumulative LF of the group
  members (lower panel). The lens magnitude is shown with an arrow.}
\end{center}
\end{figure*}

To investigate the influence of environment,
Figure~\ref{fig:cassenviron} shows $80^\prime \times 80^\prime$
grey-scale images of the number density of galaxies detected by SDSS,
together with a $8^\prime \times 8^\prime$ zoom-in on the lens. To
study the immediate neighbours of the lens galaxy, we use the SDSS
photometric redshifts, as shown in the upper right panel.  To build
the luminosity function, we select galaxies with a redshift lying
within the range $\zl \pm 0.1$, as shown in the top-right
panels. Typically for the CASSOWARY lenses, the galaxies lie in poor
groups with $\sim 20$ members, rather than rich clusters, and are the
brightest galaxies by 1 or 2 magnitudes in $r$.

\begin{table}
\caption{Observed and derived properties of the CSWA~2 and 3
  lenses. The data on CSWA~1 from Belokurov et al. (2007) is shown
  for comparison.}
\begin{tabular}{l|l|l|l}
\hline
Parameter & CSWA~1 & CSWA~2 & CSWA~3 \\
  & \tiny{Cosmic Horseshoe} & \tiny{Cheshire Cat} & \\ \hline
$\alpha_\mathrm{Lens}$ & 11:48:33.1 & 10:38:47.95 & 12:40:32.28 \\
                    &            & 10:38:39.20 &            \\
$\delta_\mathrm{Lens}$ & 19:30:03.2 & 48:49:17.9 & 45:09:02.9  \\
                    &            & 48:49:20.3 &            \\
$z_\mathrm{Lens}$      & 0.444      & 0.426      & 0.274      \\
                    &            & 0.432      &            \\
$z_\mathrm{Source}$      & 2.379      & 0.97       & 0.725      \\
                  & & $> 1.4$ & \\
$(g,i)_\mathrm{Lens}$  & 20.8,18.2       & 20.2, 17.9        & 19.6, 17.8        \\
                    &            & 20.5, 18.1        &            \\
$R_\mathrm{eff}$ & $1.7^{\prime\prime}$ & $1.7^{\prime\prime}$ &
$2.5^{\prime\prime}$ \\
                    &            & $1.9^{\prime\prime}$ &            \\
$(g,i)_\mathrm{Source}$  & 20.1,19.7        & 21.5, 20.6        & 19.7,
19.3       \\
   &         & 21.0, 20.5        &         \\
$R_\mathrm{E}$ & $5.1^{\prime\prime}$ & $7.8^{\prime\prime}$ &
$3.8^{\prime\prime}$\\
 & & $11.5^{\prime\prime}$ & \\
$M_\mathrm{E} /10^{12} \msun$& 5.4 & 17.7 & 2.4 \\ 
& & 33 & \\
\hline
\end{tabular}
\label{tab:struct}
\end{table}
\begin{figure*}
\begin{center}
\includegraphics[height=10cm]{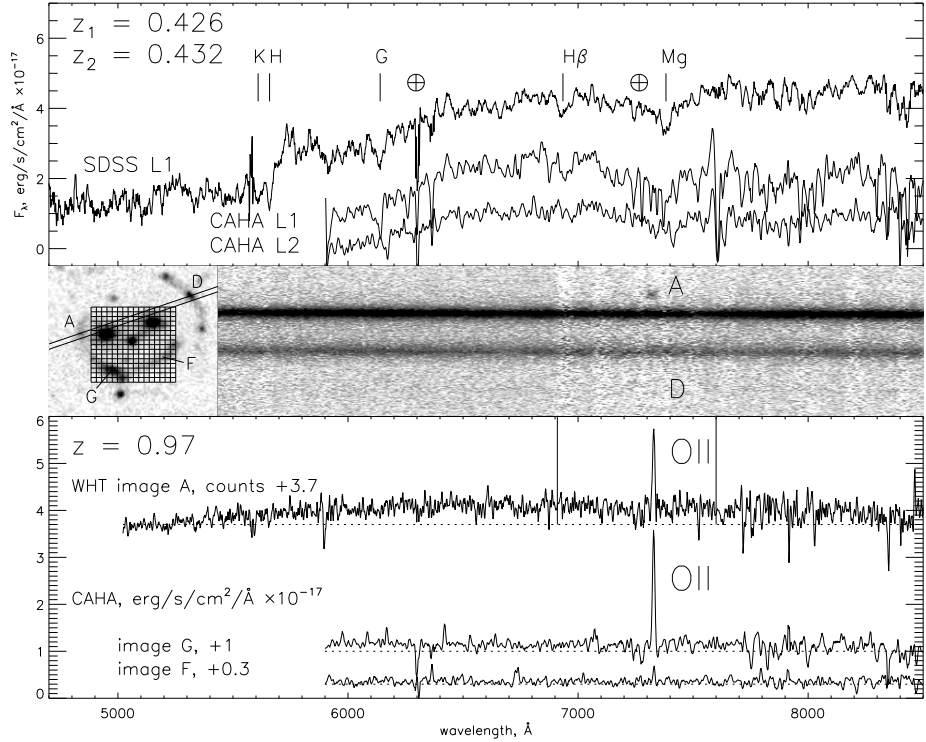}
\caption{\label{fig:catspectro} Spectroscopic data for CSWA~2 from
  SDSS, Calar Alto and the WHT. The top panel shows spectra of the two
  lens galaxies from SDSS and Calar Alto PMAS. Both galaxies exhibit
  features typical of early-type galaxies, including Ca K \& H, G-band
  and Mgb absorption lines. The two galaxies are at slightly
  different redshifts, $z_1 = 0.426$ and $z_2 = 0.432$. The $\oplus$
  symbols mark the location of prominent telluric
  absorption. The middle left panel shows the ISIS slit position and
  the PMAS IFU location. The middle right panel shows part of the ISIS
  2D spectrum with a $\sim 30^{\prime\prime}$ section, to include images
  A and D, along the slit plotted vertically and
  wavelength plotted horizontally. The lower panel shows WHT and Calar 
  Alto spectra of the
  arcs. In the PMAS data, we see prominent [O II] emission in G, but
  not in F. Similarly, in the ISIS data, the [O II] emission is clearly
  visible in the 2D cutout and the 1D extracted spectrum but
  there is no corresponding emission at the location of D. Images A, G,
  and probably B, are produced by a source at redshift $z = 0.97$,
  whilst D, F, and most likely C and E, are at a higher
  redshift.}\end{center}
\end{figure*}
\begin{figure*}
\begin{center}
\includegraphics[height=4.1cm]{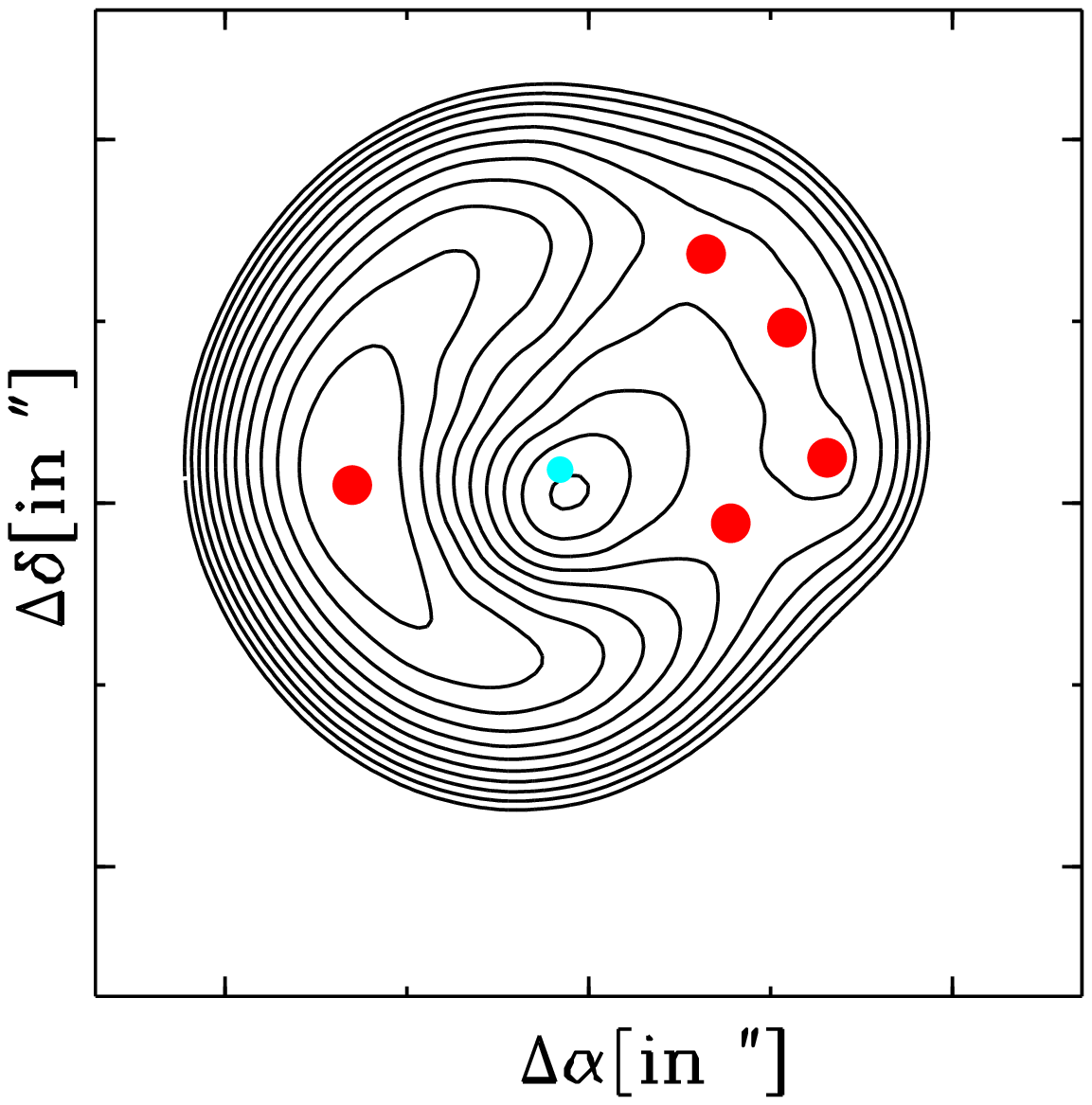}
\includegraphics[height=4.1cm]{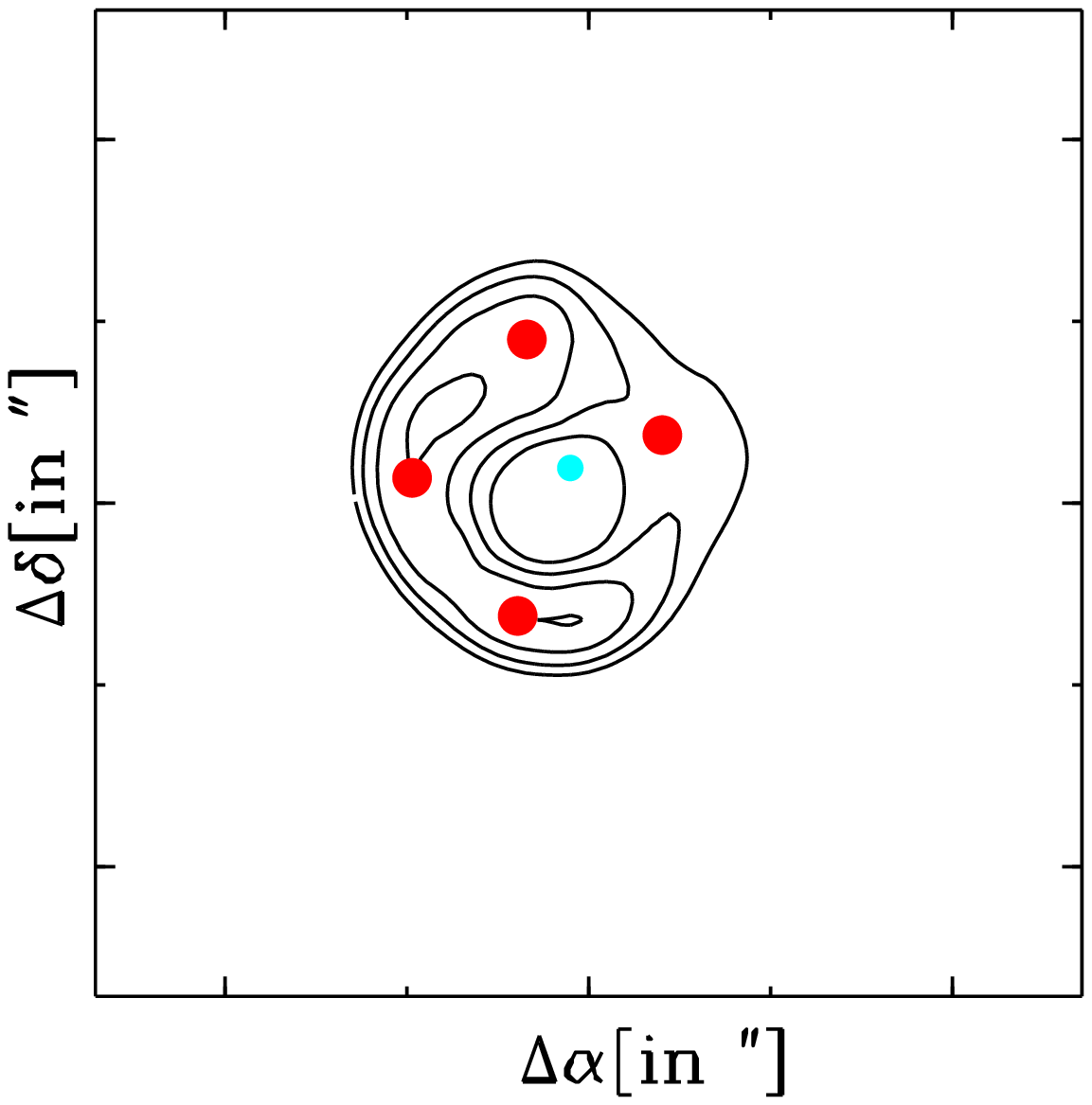}
\includegraphics[height=4.1cm]{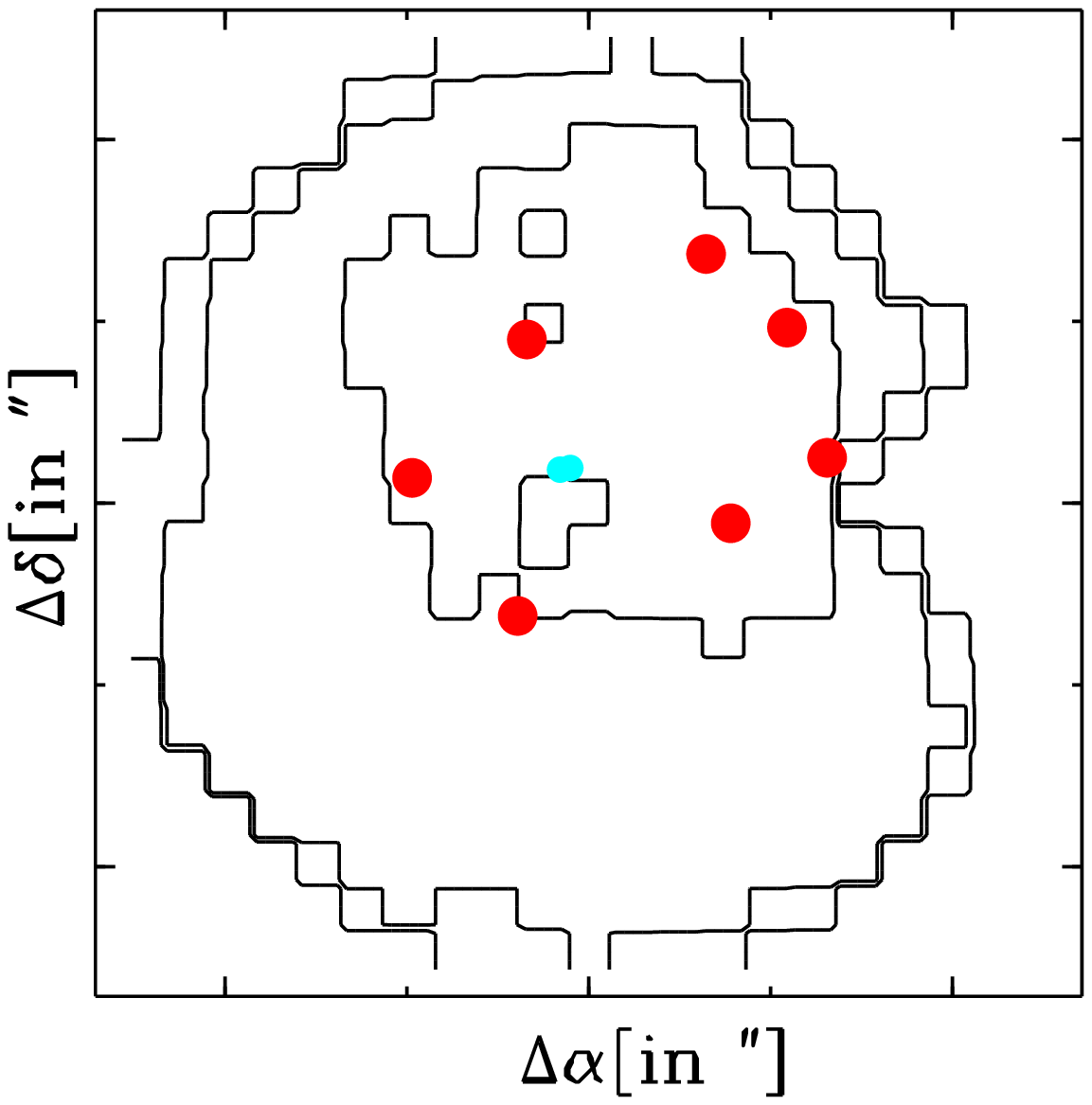}
\caption{\label{fig:cass2model} A {\tt PixeLens} model of CSWA~2 with
  image locations (red dots) and source locations (black dots). The
  left and middle panels show the Fermat surface for the $z=1.4$ and
  $z=0.97$ sources respectively, the right panel shows the pixellated
  mass distribution in units of the critical density.  Note that for
  each source additional images are predicted, which are either
  faint, or close to a bright lensing galaxy, and thus difficult to
  detect. Only the observed images are shown in the rightmost plot,
  but all the predicted images are shown in the Fermat surface plots.}
\end{center}
\end{figure*}
\begin{table*}
  \caption{Relative right ascension and declination of the images A-G 
    and the lensing galaxies L$_1$ and L$_2$ in arcseconds for CSWA~2.}
\begin{tabular}{lccccccccc}
  \hline
  \null &  L$_1$& L$_2$& A& B& C& D& E& F& G \\
  Coordinates & (4.8,1.2) & (-4.05,3.75) & (9.7,1.4) & (3.4,9.0) &
  (-6.45,13.7) & ( -10.9,9.65) & (-13.1. 2.5) & (-7.8,-1.1) & (3.9,
  -6.2)\\\hline
\end{tabular}
\label{tab:obsdata}
\end{table*}
\section{CSWA~2, The Cheshire Cat}

CSWA~2, nicknamed ``The Cheshire Cat'' (Carroll, 1866) because
of its broad smile, is a complicated system. The Cat's eyes are two
giant early-type galaxies, which are probably responsible for most of
the lensing. Our images are centered on the Cat's nose at
$\alpha = 10:38:43.10, \delta = +48:49:16.50$, which is most likely a
foreground object. The southern arc, which is the Cat's smile, is the
brightest arc in the system and is clearly visible in the SDSS
imaging. Given the complexity of CSWA~2, we obtained integral field and
long slit spectroscopy with the Calar Alto 3.5m and the WHT respectively.

Calar Alto observations were performed using the Potsdam
Multi-Aperture Spectrophotometer (PMAS, Roth et al. 2005), at the 3.5m
telescope on 2007 May 10/11. PMAS is an integral field spectrograph
that comprises a lens array with 256 individual lenses. They sample
the sky in a regular square grid of 16 $\times$ 16 elements, each with
an individual field-of-view of 1$\arcsec$ in the selected
configuration. A low-resolution grating (V300) giving a FWHM
resolution of $\sim$7\AA\ was used, covering the wavelength range
5780-9125\AA. An exposure of 3600\,s was taken on the central parts of
CSWA~2, which included both lensing galaxies and the brightest knot in
the southern arc. The data were then reduced using R3D (S\'anchez et
al. 2006). The reduction steps include bias and
pixel-to-pixel transmission correction, spectra tracing and
extraction, wavelength calibration, fibre-transmission difference
corrections and flux calibration. The frames were sky
subtracted using E3D (S\'anchez et al. 2004), by selecting the spectra
corresponding to fibres free of object signal within the
field-of-view. The resulting median sky spectrum was then subtracted
from all the fibres. Finally, the spectra were reordered
using the PMAS position table to create the reduced datacube.

A long-slit spectrum of knots A and D was obtained using the
ISIS spectrograph on the WHT on the night of 2008 May 23.  The
observations were obtained as part of the WHT Service Programme in
less than ideal conditions of high humidity, with variable seeing
around $1.6^{\prime\prime}$.  The ISIS spectrograph was configured
with the 5300\AA \ dichoic and gratings R300B (blue arm) and R158R
(red arm).  A slit-width of $1.5^{\prime\prime}$ was employed and a
total on-sky exposure time of 4500\,s was obtained, divided into five
individual exposures of 900\,s.  Wavelength calibration was performed
using exposures of standard calibration lamps. In practice, no useful
information was obtained from the blue arm, but the data from the red
arm provided wavelength coverage of $5300-10000$\,\AA\ with a
resolution of $\simeq$12\AA\ FWHM.  The data were obtained without any
on-chip binning and the spatial scale along the slit was 0.22\,\arcsec
per pixel.  Standard reduction procedures were followed using {\sc
  iraf}\footnote{{\sc iraf} is distributed by the National Optical
  Astronomy Observatories, which are operated by the Association of
  Universities for Research in Astronomy, Inc.  under cooperative
  agreement with the National Science Foundation.}  routines.  The
five individual object exposures were then combined, employing a
sigma--clipping algorithm to eliminate cosmic-rays, and the spectra of
the two central galaxies extracted.  The spectrum-trace of the
brighter galaxy was then repositioned along the slit to extract
spectra of the two arc components A and D.

Figure~\ref{fig:catspectro} summarises the available spectroscopic
data for CSWA~2. The inset in the middle panel shows the placing of
the PMAS IFU and the WHT long slit. The top panels shows the spectra
for the lensing galaxy L$_1$ (SDSS and Calar Alto) and L$_2$ (just
Calar Alto), which are typical of early-type galaxies, with Ca K \& H,
G-band, Mgb absorption, together with an indication of H$_\beta$
absorption. L$_1$ and L$_2$ lie at slightly different redshifts,
namely $z_1 = 0.426$ and $z_2 = 0.432$, which corresponds to a
Hubble-flow separation of $14.3$ Mpc in the standard cosmology
($\Omega_{\mathrm m} = 0.3, \Omega_\Lambda = 0.7, h = 0.7$.)  It is
also worth noting that L$_2$ is a radio source as detected by FIRST
with 6.9 mJy flux.

For the sources, we have spectra of knots A and D with ISIS, and F and
G with PMAS. The 2D ISIS spectra are shown in the middle panel and the
1D spectra from both ISIS and PMAS are shown in the lower panel. The
main feature is the two strips in the middle of the 2D cut-out
corresponding to the continua of the lensing galaxies. Also visible in
the range 6915 - 7600 \AA\ is the narrow emission line [O II] $\lambda
3728$ from a source at $z = 0.97$. As evident from the 2D spectra, the
emission feature is not detected at the location of the D image. The
presence/absence of [OII] emission is corroborated by the PMAS data in
the lower panels, with G exhibiting strong [O II], but not F (although
the signal from F is weak and so the evidence is not
conclusive). Given the absence of [O II] up to $\sim$ 9300 \AA\ in the
ISIS spectra, we conclude that the redshift of the second source is
likely in excess of $z = 1.4$.

As a simple and flexible tool for modelling, we take advantage of the
publicly available {\tt PixeLens}~\footnote{
  http://www.qgd.uzh.ch/projects/pixelens/} code~\citep{Sa04}. Here,
the mass distribution is pixellated into tiles. It is easy to find
many possible mass tilings that reproduce the positions of the images
exactly. {\tt PixeLens} overcomes this problem by restricting
attention to mass distributions for which (i) the density gradient
anywhere must point within $45^\circ$ of the centre, (ii) the
projected density of any pixel must be less than twice the average
value of its neighbors and (ii) the surface density radial profile
$\kappa(r)$ must be steeper than $r^{-0.5}$. These constraints
guarantee that the model is centrally concentrated, smooth and roughly
isothermal $\kappa \sim r^{-1}$.  

{\tt PixeLens} samples the solution space using a Markov chain Monte
Carlo method, typically generating an ensemble of 100 models that
reproduce the input data, which in our case are the image locations
and parities. As the constraint equations are linear, averaging the
ensemble also produces a solution. All figures and numbers given from
the modelling are ensemble-averages over a set of 100 models. Note
that our models always reproduce the relative positions of the lensing
galaxies and the images exactly, but they usually predict fainter
additional images. The quality of our available images is not adequate
to allow useful quantitative measurements of the brightnesses of the
individual components. Nonetheless, in our modelling, we give the
magnifications of the images to show that there is a qualitative match
between the predictions and the data.

A {\tt PixeLens} model for CSWA~2 using the input data in
Table~\ref{tab:obsdata} is shown in Figure~\ref{fig:cass2model}. The
left and middle panels show the Fermat surfaces for the sources at
$z=1.4$ and $z = 0.97$ respectively.  The positions of the images of
the higher redshift source, C,D, E and F, are all reproduced. The
total magnification is $\sim 45 \pm 7$, with C and D as the brightest
images (magnifications of $15.9$ and $14.2$) followed by E and F
(magnifications of $7.1$ and $5.1$). There is also a predicted faint
counterimage at $\Delta \alpha \approx -13^{\prime\prime}, \Delta
\delta \approx 1^{\prime\prime}$ with a magnification of $3.1$ and, of
course, a highly demagnified central image.  For the lower redshift
source, the image locations of G, A and B are reproduced. The
predicted respective magnifications are $3.9, 3.6$ and $1.3$, which is
in reasonable accord with the relative brightness of the images in
Figure~\ref{fig:cassdata}. There is a bright counterimage with a
magnification of $4.3$ offset from the centre by $\Delta \alpha
\approx -4^{\prime\prime}, \Delta \delta \approx
4^{\prime\prime}$. The predicted counterimage is close to the location
of the lensing galaxy L$_2$, which may render its detection
difficult. The total magnification of the $z = 0.97$ source is $\sim
14 \pm 4$. Note that the density contours are roundish but with
evident substructure, whilst the mass enclosed $M_\mathrm{E}$ within
the outer ring of radius $11.5^{\prime\prime}$ is a stupendous $33
\times 10^{12}\ \msun$ -- almost an order of magnitude greater than
the mass in the entire Local Group!

\begin{figure*}
\begin{center}
\includegraphics[height=9cm]{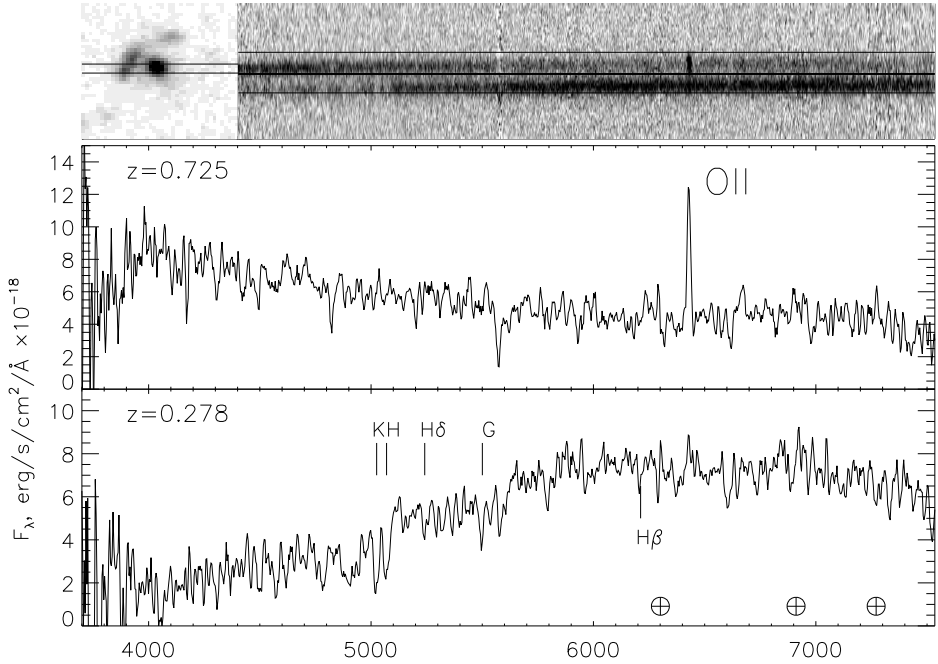}
\caption{\label{fig:cass3spectrum} Spectroscopic follow-up of
  CSWA~3. The top panel shows the slit position and the 2D spectrum
  from the SAO SCORPIO Instrument. The 1D spectra are extracted from
  the rectangular regions indicated and shown in the lower two panels. For the
  source, there is an unambiguous detection of the [O II] line at $z =
  0.725$. The lens spectra is typical of an early-type galaxy with Ca
  K \& H lines clearly visible and a redshift $ z =
  0.278$. The $\oplus$ symbols mark the location of prominent
  telluric absorption features.}\end{center}
\end{figure*}

\section{CSWA~3}

Long-slit spectral observations of CSWA~3 were performed on 2008 March
28/29 with the multi-mode focal reducer SCORPIO
\citep{2005AstL...31..194A} installed at the prime focus of the
BTA~6-m telescope at the SAO in $1.1^{\prime\prime}$ seeing.  The slit
was placed to include some of the light from the lens galaxy, as shown
in the inset of Figure~\ref{fig:cass3spectrum}.  We used the VPHG550G
grism which covers the wavelength interval 3650--7550\,\AA\ with a
spectral resolution 8-10\,\AA\ FWHM.  With a CCD EEV 42-40
2k\,$\times$\,2k detector, the reciprocal dispersion was $1.9$\,\AA\
per pixel.  The total exposure time was 1800\,s, divided into two 900
s exposures. The target was moved along the slit between exposures to
ease background subtraction and CCD fringe removal.  The bias
subtraction, geometrical corrections, flat fielding, sky subtraction,
and calibration to flux units ($F_{\lambda}$) was performed by means
of IDL-based software. From its spectrum, the source is a typical
star-forming galaxy, with a redshift of $z = 0.725$, estimated from
the prominent [O II] $\lambda$3728 emission line. There are no other
viable redshift identifications given the lack of additional emission
features in the spectrum. We also measured the early-type galaxy lens 
redshift, $z = 0.278$, which was not included in the SDSS. 

\begin{figure}
\begin{center}
\includegraphics[height=4.1cm]{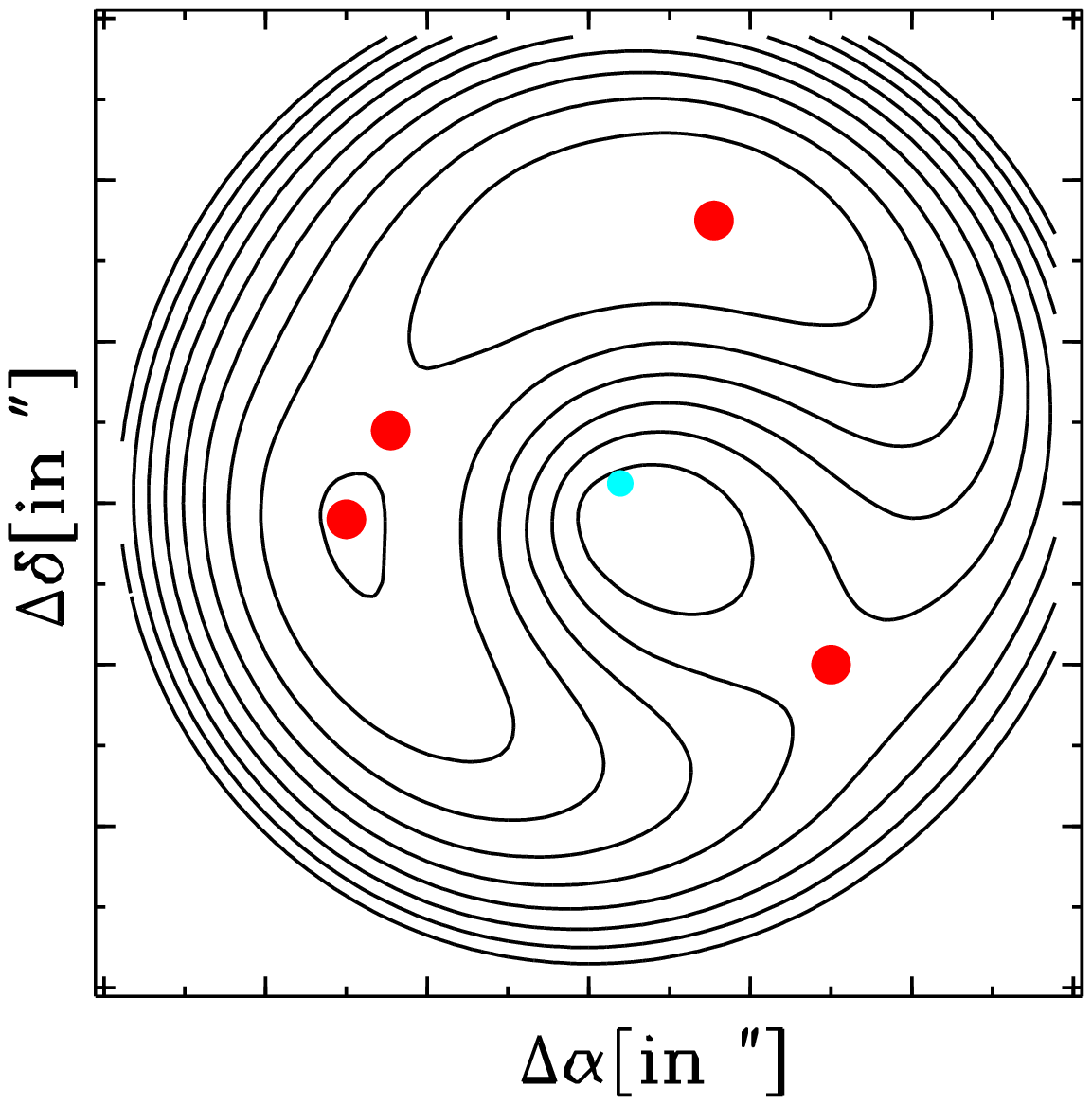}
\includegraphics[height=4.1cm]{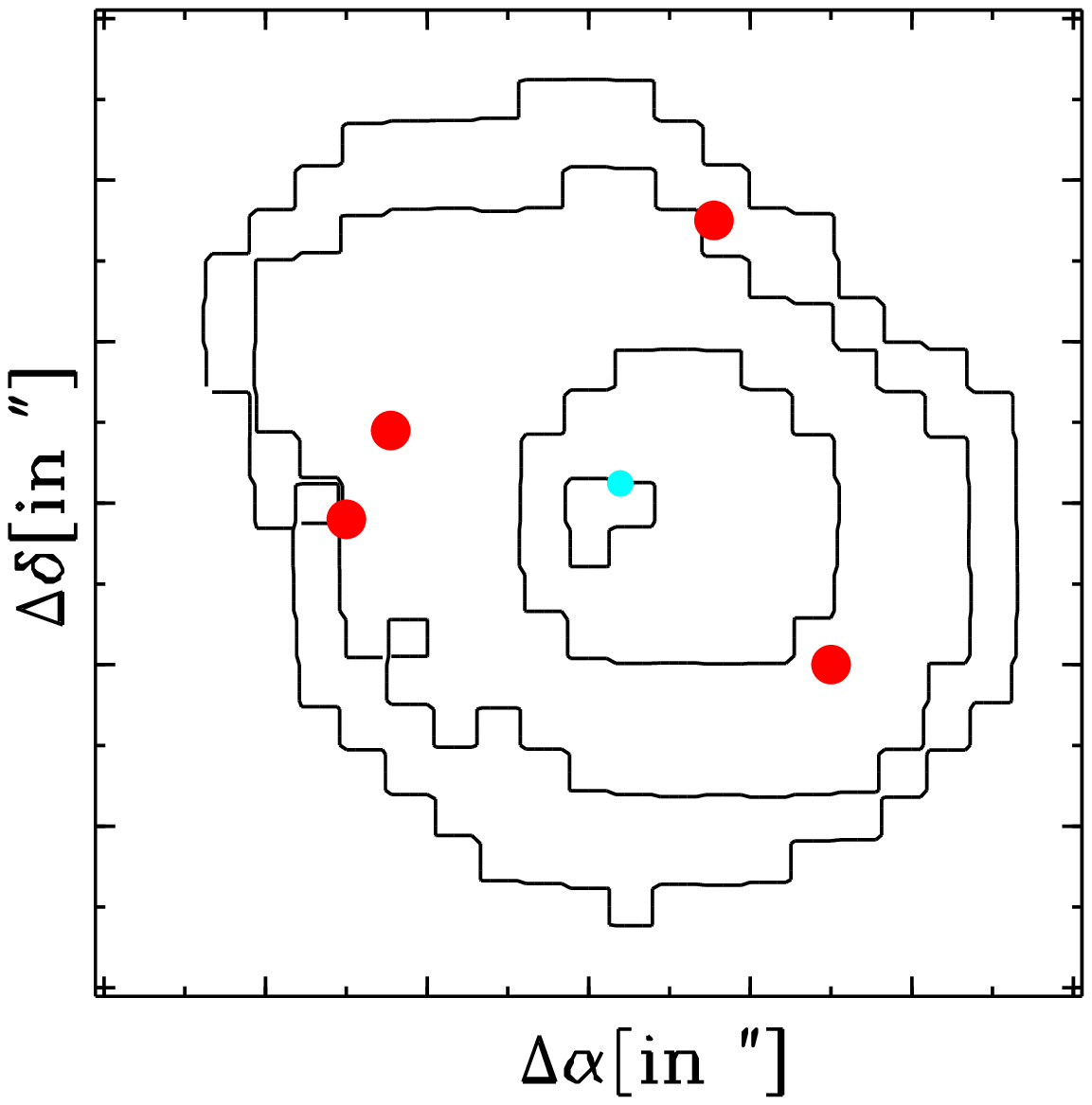}
\caption{\label{fig:cass3model} A {\tt PixeLens} model of the CSWA~3
  lens. The left panel shows the Fermat surface, the right panel the
  pixellated mass distribution in units of the critical density. The
  image locations are shown as red dots, whilst the source location is
  shown as a blue dot. Note that there is a predicted fourth image at
  the saddle of the Fermat surface (as well as of course a demagnified
  central image).}
\end{center}
\end{figure}
\begin{table}
  \caption{Relative right ascension and declination of the 
    images A-C and the lensing galaxy L in arcseconds for CSWA~3.}
\begin{tabular}{lcccc}
  \hline
  \null & L & A & B & C \\
  Coordinates & (0.00,0.00) & (3.00,-0.20) & (2.45,0.90) & (-1.55, 3.50)\\\hline
\end{tabular}
\label{tab:obsdatacass3}
\end{table}

CSWA~3 lies in a known galaxy cluster, namely NSCS J124034+450923, as
identified in the DPOSS survey by \citet{Lo04}. The cluster itself is
a ROSAT detected X-ray source. CSWA~3 is probably a quadruplet, with
the faintest image undetected. To test this hypothesis, we build a {\tt
  PixeLens} model, using the data on image locations given in
Table~\ref{tab:obsdatacass3}. An ensemble-averaged solution for the Fermat
surface and the projected mass distribution is shown in the left and
right panels of Figure~\ref{fig:cass3model}.  The arc (or at least
what we see of it) is quite circular, hinting that shear from the
galaxy cluster plays a modest role. The lensing mass within the arc is
$\sim 2.5 \times 10^{12}\ \msun$ -- consistent with most of the
lensing effect coming from a single massive early-type galaxy.  Note
that the contours of surface mass in the model are aligned in roughly
the same direction as the luminous material in the lensing galaxy.

The {\tt PixeLEns} model allows us to estimate the total magnification
as $\sim 18 \pm 3$. B is the brightest image with a magnification of
$\sim 7.6$, followed by A (magnification of $\sim 4.2$) and then C
(magnification of $\sim 3.5$). The model predicts a fourth, fainter
image D with magnification of $\sim 2.5$, at the location of the
saddle-point of the limacon in the Fermat time delay surface.  In the
model, its predicted location is offset by $\Delta \alpha \approx
-3^{\prime\prime}, \Delta \delta \approx -2.0^{\prime\prime}$ from the
lens galaxy centre. This missing image should be detectable with
deeper imaging. There is of course also a fifth, demagnified
central image.

\section{Conclusions}

We have described {\it The CAmbridge Sloan Survey Of Wide ARcs in the
  skY} (CASSOWARY). This is a search for multiple blue components
around massive red galaxies using data from the SDSS.
The search strategy is simple and fast, enabling many
experiments with different cuts to be performed quickly, generating
many possible lens candidates. A ranked list of our best candidates is
available at ``http://www.ast.cam.ac.uk/research/cassowary/''
It is clear from the recent discoveries by the CASSOWARY
group~\citep{Be07} and others~\citep{Al07, Sh08} that there are many
such lenses within the SDSS.

We have presented new imaging, spectroscopy and modelling for two of our
wide separation gravitational lenses, CSWA~2 and CSWA~3.  These
two lenses are distinguished from among many candidates only by the
fact that we have obtained detailed spectroscopic and
photometric follow-up.

CSWA~2, ``the Cheshire Cat'', is a complex system in which there are 
two lensing galaxies and at least two sources. The Cat's eyes are the lenses,
namely two massive early-type galaxies at $z = 0.426$ and $0.432$
respectively, whilst the Cat's nose is a foreground galaxy that
has minimal lensing effect. The two sources are a star-forming galaxy
at $z = 0.97$ and a high redshift galaxy ($z> 1.4$). They are
transmogrified by gravitational lensing into giant arcs which form the
Cat's smile and eyebrows. The mass enclosed within the larger arc of
radius $\sim 11^{\prime\prime}$ is $\sim 33 \times 10^{12}\ \msun$,
greater than the Local Group! 

CSWA~3 is simpler and has a standard
morphology of three bright images on one side of the lensing galaxy and
one faint image on the other side, although the latter is not detected in
the SDSS data. The source is a star-forming galaxy at $z = 0.725$,
which is being lensed by a foreground elliptical at $z = 0.274$. The
radius of the arc is $\sim 4^{\prime\prime}$ and the enclosed mass is
$\sim 2.5 \times 10^{12}\ \msun$, typical for a giant elliptical.

The study of large-separation lenses is important as a probe of the
halo mass function. The extent and density profile of dark haloes, as
well as possible signatures of the compression of dark halos by
cooling baryons, may be extracted from the distribution of lens
separations~\citep{Ko01}. The regime intermediate between the
arcsecond separation lenses of typical strong galaxy lensing and
cluster lensing is therefore of strategic significance in modern
astrophysics.  Surveys such as CASSOWARY and its
competitors~\citep{Ku07,Sh08} are providing our first forays into this
new terrain.

\section*{Acknowledgements}

We thank the referee, Peter Schneider, for many helpful suggestions
that have improved the paper. NWE, PCH and RGM acknowledge support
from the STFC-funded Galaxy Formation and Evolution programme at the
Institute of Astronomy.  Funding for the SDSS and SDSS-II has been
provided by the Alfred P.  Sloan Foundation, the Participating
Institutions, the National Science Foundation, the U.S. Department of
Energy, the National Aeronautics and Space Administration, the
Japanese Monbukagakusho, the Max Planck Society, and the Higher
Education Funding Council for England. The SDSS Web Site is
http://www.sdss.org/.  We thank Pablo Rodriguez-Gil for undertaking
the long-slit observations at the WHT.  The WHT and its service
programme are operated on the island of La Palma by the Isaac Newton
Group in the Spanish Observatorio del Roque de los Muchachos of the
Instituto de Astrofsica de Canarias. The paper is also based on
observations collected at the Centro Astron\'mico Hispano Alem\'an
(CAHA) at Calar Alto, operated jointly by the Max-Planck Institut
f\"ur Astronomie and the Instituto de Astrof\'\i sica de Andaluc\'\i a
(CSIC).  The paper was partly based on observations collected with the
6m telescope of the Special Astrophysical Observatory (SAO) of the
Russian Academy of Sciences (RAS) which is operated under the
financial support of Science Department of Russia (registration number
01-43).  A.V.M. also acknowledges a grant from the President of
Russian Federation (MK1310.2007.2)"

\end{document}